\documentclass[preprint,showpacs,preprintnumbers,amsmath,amssymb]{revtex4}

\usepackage{graphicx}% Include figure files
\usepackage{dcolumn}% Align table columns on decimal point
\usepackage{bm}% bold math
\usepackage{latexsym}
\usepackage{amsfonts}
\usepackage{amssymb}
\usepackage{amsmath}
\usepackage{color}
\begin{document}

\preprint{KUNS 2306}

\title{Anisotropic Power-law Inflation }

\author{Sugumi Kanno$^{1)}$}
\author{Jiro Soda$^{2)}$}
\author{Masa-aki Watanabe$^{2)}$}
\affiliation{1) Institute of Cosmology, Department of Physics and Astronomy, 
Tufts University, Medford, Massachusetts 02155, USA}
\affiliation{2) Department of Physics,  Kyoto University, Kyoto, 606-8501, 
Japan}

\date{\today}% It is always \today, today,
             %  but any date may be explicitly specified

%===============================================================%
%************************* ABSTRACT ****************************%
%===============================================================%
\begin{abstract}
We study an inflationary scenario in supergravity model 
with a gauge kinetic function.
We find exact anisotropic power-law inflationary solutions 
when both the potential function for an inflaton and the gauge kinetic function
are exponential type. The dynamical system analysis tells us 
that the anisotropic power-law inflation is an attractor for a large parameter 
region. 
\end{abstract}

% PACS, the Physics and Astronomy
                             % Classification Scheme.
%\keywords{Suggested keywords}%Use showkeys class option if keyword
                              %display desired
\maketitle

%===============================================================%
%************************ SECTION I ****************************%
%===============================================================%
\section{Introduction}

It is believed that inflationary scenario explains main features 
of cosmic microwave background radiation (CMB) observed by WMAP~\cite{Komatsu:2010fb}. 
The remarkable predictive power of inflation is associated with the cosmic
no-hair conjecture which claims that any classical hair will disappear
exponentially fast once the vacuum energy dominates the universe.
As a result, there remain quantum vacuum fluctuations which become
seeds of the large scale structure of the universe. 
The quantum vacuum fluctuations give rise to the following robust predictions. 
The power spectrum of fluctuations should have the statistical isotropy 
 due to the symmetry of deSitter background. 
 The scale dependence of power spectrum should be absent because of the time translation
invariance of deSitter spacetime. The slow roll conditions should suppress nonlinearity
and hence yield Gaussian statistics of fluctuations. 
These predictions have been observationary proved in rough.

Nowadays, however, precise cosmological observations force us to look at fine
structures of an inflationary universe. Actually, the inflationary universe is not
exactly deSitter and deviation from the exact deSitter is characterized
by slow roll parameters of the order of a few percent. 
Indeed, the deviation from the scale free spectrum
is well known. Moreover, it has been shown that the deviation from Gaussianity
in a single inflaton model is determined by the slow roll
 parameters~\cite{Maldacena:2002vr}. From these examples of deviations,
it would be quite natural to expect statistical anisotropy of the order of 
the slow roll parameter~\cite{Yokoyama:2008xw,Dimopoulos:2009vu,ValenzuelaToledo:2009af,ValenzuelaToledo:2009nq,Dimastrogiovanni:2010sm,BlancoPillado:2010uw,
Adamek:2010sg}. 
 However, a prejudice due to the cosmic no-hair conjecture has prevented us seeking
 the statistical anisotropy seriously.
 
 Historically, there have been challenges to the cosmic no-hair 
 conjecture~\cite{Ford:1989me,Kaloper:1991rw,Barrow:2005qv,Barrow:2009gx,Campanelli:2009tk,Golovnev:2008cf,Kanno:2008gn,Ackerman:2007nb}. 
Unfortunately, it turned out that these models 
suffer from either the instability~\cite{Himmetoglu:2008zp,Golovnev:2009rm,EspositoFarese:2009aj}, or a fine tuning problem, or a naturalness problem. 
Recently, however, we have succeeded in finding stable anisotropic inflationary solutions in a natural set-up,
which provide counter examples to the cosmic no-hair 
conjecture~\cite{Watanabe:2009ct,Kanno:2009ei}. More precisely, 
we have shown that, in the presence of a vector field coupled with the inflaton,
there could be a small anisotropy in the expansion rate which never
decays during inflation. Furthermore, since the anisotropic inflation is an attractor
solution, the predictive power of the model still remains~\cite{Himmetoglu:2009mk,Dulaney:2010sq,Gumrukcuoglu:2010yc,Watanabe:2010fh}.
 The imprints of the anisotropic expansion could be
 found in the fine structures of the CMB~\cite{Watanabe:2010bu}.

Our anisotropic inflationary model is motivated by supergravity which
can be regarded as a low energy limit of superstring theory.
It is well known that the supergravity models can be constrained
 by comparing predictions of inflation with cosmological observations.
 For example, the tilt of the spectrum gives interesting information
of the superpotential $W(\phi^i)$ and the Kaler potential $K(\phi^i,\bar{\phi}^i)$
which are functionals of complex scalar fields. Here, a bar represents
a complex conjugate. The information will provide a hint to the fundamental theory.
More concretely, the bosonic part of the action of supergravity reads
\begin{eqnarray}
S = \int d^4 x \sqrt{-g}
\left[ \frac{1}{2\kappa^2}R 
+ K_{i\bar{j}} \partial^\mu \phi^i \partial_\mu \bar{\phi}^{\bar{j}} 
   - e^K K^{i\bar{j}} D_i W D_{\bar{j}} \bar{W} 
   - f_{ab}(\phi) F^a_{\mu\nu} F^{b\mu\nu} \right]
\end{eqnarray}
where we have defined
$K_{i\bar{j}} = \partial^2 K/\partial \phi^i \partial \bar{\phi}^j$
and $D_i W=\partial W/\partial \phi^i + \partial K/\partial \phi^i W$. 
Here, we note that the gauge kinetic function $f_{ab}(\phi)$ in front of the vector
kinetic term $F^a_{\mu\nu} F^{b\mu\nu}$ 
could be nontrivial functions of the scalar fields.
Interestingly, so far, the vector part has been neglected in most of 
discussion of inflation. 
The reason is partially due to the cosmic no-hair conjecture
which states that the anisotropy, curvature, and any matter will vanish
once the inflation commences. As we mentioned, however
this is not true when we look at percent level fine structures of
inflationary scenarios. 

From the point of view of precision cosmology,
it is important to explore the role of the gauge kinetic function
in inflation. In the previous work, we have used rather unconventional
functional form $\propto \exp(c \phi^2) $ (c is a constant) 
for the gauge kinetic function~\cite{Martin:2007ue}. 
Although we have also mentioned general
cases, it is desirable to investigate more natural cases explicitly.
It is well known that the exponential type functional form
is ubiquitous from the point of view of dimensional reduction of
higher dimensional theory. Hence, in this paper, we study the cases
where a functional form for the potential and the gauge kinetic function
is an exponential dependence on the inflaton. In particular, 
we present exact anisotropic inflationary solutions.
Moreover, we demonstrate these solutions become attractors for a 
large parameter region. 

The organization of the paper is as follows.
In section II, we introduce inflationary models where the vector
field couples with an inflaton and obtain exact solutions which contain
anisotropic power-law solutions. 
In section III, we investigate solution space with dynamical system 
approach and show that the anisotropic inflation
is an attractor for a large parameter region.
 The final section is devoted to the conclusion.

%===============================================================%
%************************ SECTION II ****************************%
%===============================================================%
\section{Exact Anisotropic Inflationary Solutions}
\label{sc:basic}

In this section, we consider a simple model with exponential
potential and gauge kinetic functions and then find exact power-law solutions.
In addition to a well known isotropic power-law solution, we find 
an anisotropic power-law inflation. 

We consider the following action for the gravitational field, the inflaton
 field $\phi$ and the
vector field $A_\mu$ coupled with $\phi$:
\begin{eqnarray}
S=\int d^4x\sqrt{-g}\left[~\frac{M_p^2}{2}R
-\frac{1}{2}\left(\partial_\mu\phi\right)\left(\partial^{\mu}\phi\right)
-V(\phi)-\frac{1}{4} f^2 (\phi) F_{\mu\nu}F^{\mu\nu}  
~\right] \ ,
\label{action1}
\end{eqnarray}
where $g$ is the determinant of the metric, $R$ is the
Ricci scalar, respectively.
Here, $M_p $ represents the reduced Planck mass.
 The field strength of the vector field is defined by 
$F_{\mu\nu}=\partial_\mu A_\nu -\partial_\nu A_\mu$. 
Motivated by the dimensional reduction of
higher dimensional theory such as string theory, we assume the exponential potential
\begin{eqnarray}
V (\phi) = V_0 e^{\lambda \frac{\phi}{ M_p}}  
\end{eqnarray}
and the exponential gauge kinetic function
\begin{eqnarray}
f(\phi) = f_0 e^{\rho \frac{\phi}{M_p} } \ .
\end{eqnarray}
In principle, the parameters $V_0 , f_0 , \lambda$, and $\rho$ 
can be determined once the compactification scheme is specified.
However, we leave those free in this paper. 

Using the gauge invariance, we can choose the gauge $A_0 =0$.
Without loosing the generality, we can take the $x$-axis in the direction
of the vector field. Hence, we have the homogeneous fields of the form
$
A_\mu=(~0,~v(t),~0,~0~)
$
and 
$ 
\phi=\phi(t) \ .
$
As there exists the rotational symmetry in the $y$-$z$ plane now,
we take the metric to be 
\begin{eqnarray}
ds^2=- dt^2+e^{2\alpha(t)}\left[~ 
e^{-4\sigma(t)}dx^2    
+e^{2\sigma(t)}\left( dy^2 + dz^2\right)~\right] \ ,
\label{metric}
\end{eqnarray}
where the cosmic time $t$ is used.
Here, $e^\alpha$ is an isotropic scale factor and $\sigma$ represents
a deviation from the isotropy.  With above ansatz, 
the equations of motion can be written down as
\begin{eqnarray}
&& \dot{\alpha}^2  = \dot{\sigma}^2
+\frac{1}{3M_p^2}\left[ \frac{1}{2} \dot{\phi}^2+V(\phi)
+\frac{1}{2}f^{2} (\phi) e^{-2\alpha+4\sigma } \dot{v}^2  \right] \ , 
\label{hamiltonian}\\
&& \ddot{\alpha} = -3\dot{\alpha}^2 + \frac{1}{M_p^2} V(\phi )
 +\frac{1}{6M_p^2 }f^{2}(\phi )e^{-2\alpha +4\sigma} \dot{v}^2 \ , 
\label{evolution:alpha}\\
&& \ddot{\sigma} = -3\dot{\alpha}\dot{\sigma} 
+ \frac{1}{3M_p^2}f^{2}(\phi )e^{-2\alpha +4\sigma} \dot{v}^2
\label{eq:sigma}, \\
&& \ddot{\phi} = -3\dot{\alpha}\dot{\phi} -V'(\phi ) 
+  f(\phi ) f'(\phi ) e^{-2\alpha +4\sigma } \dot{v}^2
\label{eq:phi} \ , \\
&& \frac{d}{dt} \left[f^2 e^{\alpha +4\sigma} \dot{v} \right] =0 \ ,
\label{eq:v}
\end{eqnarray}
where an overdot and a prime denote the derivative with respect to the 
cosmic time $t$ and $\phi$, respectively.
We find Eq.~(\ref{eq:v}) can be easily solved as
\begin{eqnarray}
\dot{v} = f^{-2}(\phi ) e^{-\alpha -4\sigma}p_{A}, 
\label{eq:Ax}
\end{eqnarray}
where $p_A$ denotes a constant of integration. 
Substituting (\ref{eq:Ax}) into other equations, we obtain equations
\begin{eqnarray}
\dot{\alpha}^2  &=& \dot{\sigma}^2
+\frac{1}{3M_p^2}\left[ \frac{1}{2} \dot{\phi}^2+V(\phi)
+\frac{p_{A}^2}{2}f^{-2} (\phi) e^{-4\alpha-4\sigma }  \right] \ , 
\label{hamiltonian:0}\\
\ddot{\alpha} &=& -3\dot{\alpha}^2 + \frac{1}{M_p^2} V(\phi )
 +\frac{p_{A}^2}{6M_p^2}f^{-2}(\phi )e^{-4\alpha -4\sigma}, 
\label{evolution:alpha:0}\\
\ddot{\sigma} &=& -3\dot{\alpha}\dot{\sigma} 
+ \frac{p_{A}^2}{3M_p^2}f^{-2}(\phi )e^{-4\alpha -4\sigma} 
\label{eq:sigma:0}, \\
\ddot{\phi} &=& -3\dot{\alpha}\dot{\phi} -V'(\phi ) 
+ p_{A}^2 f^{-3}(\phi )f'(\phi ) e^{-4\alpha -4\sigma } 
\label{eq:phi:0} \ .
\end{eqnarray}

In the case of no vector fields, there exists the power-law inflationary
solution~\cite{Abbott:1984fp,Lucchin:1984yf,Barrow:1987ia,Halliwell:1986ja,Burd:1988ss}.
Therefore, let us first seek the power-law solutions by assuming
\begin{eqnarray}
  \alpha = \zeta \log t \ , \hspace{1cm}
  \sigma = \eta \log t \ , \hspace{1cm}
  \frac{\phi}{M_p} = \xi \log t + \phi_0 \ .
\end{eqnarray}
Apparently, for a trivial vector field $p_A =0$,
we have the isotropic power-law solution
\begin{eqnarray}
  \zeta = \frac{2}{\lambda^2} \ , \hspace{1cm} \eta=0 \ , \hspace{1cm} 
  \xi = - \frac{2}{\lambda} \ , \hspace{1cm} 
  \frac{V_0}{M^2_p} e^{\lambda \phi_0} = \frac{2(6-\lambda^2)}{\lambda^4} \ .
  \label{iso-power}
\end{eqnarray}
In this case, we have the spacetime
\begin{eqnarray}
 ds^2 = -dt^2 + t^{4/\lambda^2} \left( dx^2 +dy^2 + dz^2 \right) \ .
 \label{isotropic}
\end{eqnarray}
Hence, we need $\lambda \ll 1$ for obtaining the sufficiently fast expansion. 

Next, interestingly, we see that there exists the other non-trivial exact 
solution in spite of the existence of the no-hair theorem~
\cite{Wald:1983ky,Kitada:1991ih,Kitada:1992uh}.
From the hamiltonian constraint equation (\ref{hamiltonian:0}),
we set two relations
\begin{eqnarray}
  \lambda \xi = -2  \ , \hspace{1cm} \rho \xi +2 \zeta + 2\eta =1
  \label{A}
\end{eqnarray}
to have the same time dependence for each term. 
The latter relation is necessary 
only in the non-trivial vector case, $p_A \neq 0$.
Then, for the amplitudes to balance, we need 
\begin{eqnarray}
  -\zeta^2 +\eta^2 +\frac{1}{6} \xi^2 + \frac{1}{3} u +\frac{1}{6} w =0 \ ,
  \label{B}
\end{eqnarray}
where we have defined
\begin{eqnarray}
   u = \frac{V_0}{M_p^2} e^{\lambda \phi_0} \  ,  \hspace{1cm} 
   w = \frac{p_A^2 }{M_p^2} f_0^{-2} e^{-2\rho \phi_0}\,.
\end{eqnarray}
The equation for the scale factor (\ref{evolution:alpha:0}) 
under Eq.~(\ref{A}) yields
\begin{eqnarray}
  -\zeta + 3\zeta^2 -u - \frac{1}{6} w =0 \ .
  \label{C}
\end{eqnarray}
Similarly, the equation for the anisotropy (\ref{eq:sigma:0}) gives 
\begin{eqnarray}
  -\eta + 3 \zeta\eta - \frac{1}{3} w =0 \ .
  \label{D}
\end{eqnarray}
Finally, from the equation for the scalar (\ref{eq:phi:0}),
we obtain
\begin{eqnarray}
  -\xi + 3\zeta \xi + \lambda u -\rho w = 0  \ .
  \label{E}
\end{eqnarray}
Using Eqs.~(\ref{A}),~(\ref{C}) and (\ref{D}), we can solve $u$ and $w$ as
\begin{eqnarray}
u = \frac{9}{2} \zeta^2 - \frac{9}{4} \zeta - \frac{3\rho}{2\lambda} \zeta
                 + \frac{1}{4} + \frac{\rho}{2\lambda}  \ , \hspace{1cm}
w = -9 \zeta^2 + \frac{15}{2} \zeta + \frac{9\rho}{\lambda} \zeta 
                 -\frac{3}{2} -\frac{3\rho}{\lambda} \,.
                 \label{uw}
\end{eqnarray}
Substituting these results into Eq.~(\ref{E}), we obtain
\begin{eqnarray}
  \left( 3\zeta-1 \right) \left[ 6 \lambda \left( \lambda + 2\rho \right)\zeta 
  - \left( \lambda^2 + 8\rho \lambda + 12 \rho^2 + 8 \right) \right] =0 \ .
\end{eqnarray}

In the case of $\zeta=1/3$, we have $u=w=0$. Hence, it is not our desired 
solution. Thus, we have to choose
\begin{eqnarray}
 \zeta = \frac{\lambda^2 + 8 \rho \lambda + 12 \rho^2 +8}{6\lambda (\lambda + 2\rho)} \ .
\end{eqnarray}
Substituting this result into Eq.~(\ref{C}), we obtain
\begin{eqnarray}
 \eta = \frac{\lambda^2 + 2\rho \lambda -4 }{3\lambda (\lambda + 2\rho)}\,. 
\end{eqnarray}
From Eq.~(\ref{A}), we have
\begin{eqnarray}
 \xi = - \frac{2}{\lambda}\,.
\end{eqnarray}
Finally, Eq.~(\ref{uw}) reduce to
\begin{eqnarray}
 u = \frac{(\rho \lambda + 2\rho^2 +2)(-\lambda^2 + 4\rho \lambda +12 \rho^2 +8)}
      {2\lambda^2 (\lambda +2\rho )^2 }
\end{eqnarray}
and
\begin{eqnarray}
 w = \frac{(\lambda^2 + 2\rho \lambda -4)(-\lambda^2 + 4\rho \lambda +12 \rho^2 +8)}
      {2\lambda^2 (\lambda +2\rho )^2 } \ .
\end{eqnarray}
Note that Eq.~(\ref{B}) is automatically satisfied.

In order to have inflation, we need $\lambda \ll 1$.
For these cases, $u$ is always positive.
Since $w$ should be also positive, we have the condition
\begin{eqnarray}
 \lambda^2 + 2\rho \lambda > 4 \ .
\end{eqnarray}
Hence, $\rho $ must be much larger than one. 
Now, the spacetime reads
\begin{eqnarray}
  ds^2 = -dt^2 + t^{2\zeta-4\eta} dx^2 + t^{2\zeta +2\eta} 
\left( dy^2 + dz^2 \right) \ .
  \label{anisotropic}
\end{eqnarray}
The average expansion rate is determined by $\zeta$ and the average slow roll parameter
is given by
\begin{eqnarray}
\epsilon \equiv -\frac{\dot{H}}{H^2}
= \frac{6\lambda (\lambda + 2\rho)}{\lambda^2 + 8 \rho \lambda + 12 \rho^2 +8}
\ ,
\label{epsilon}
\end{eqnarray} 
where we have defined $H=\dot{\alpha}$. 
In the limit  $\lambda \ll 1$ and $\rho \gg 1$, 
this reduces to $\epsilon = \lambda /\rho$.  
Now, the anisotropy is characterized by
\begin{eqnarray}
  \frac{\Sigma}{H} \equiv \frac{\dot{\sigma}}{\dot{\alpha}}
  = \frac{2(\lambda^2 + 2\rho \lambda -4) }
  {\lambda^2 + 8 \rho \lambda + 12 \rho^2 +8} \ . 
\label{soverh}
\end{eqnarray}
From Eq.~(\ref{epsilon}) and (\ref{soverh}), we obtain a relation
\begin{eqnarray}
\frac{\Sigma}{H} = \frac{1}{3}I\epsilon \,,
\hspace{1cm}
I=\frac{\lambda^2 + 2\rho\lambda - 4}{\lambda^2 + 2\rho\lambda}\,.
\end{eqnarray}
In order to see the relation to  our previous result~
\cite{Watanabe:2009ct,Watanabe:2010fh}, we can rewrite the above result 
as
\begin{eqnarray}
I=\frac{c-1}{c}\,,
\hspace{1cm}
c=\frac{\lambda^2 + 2\rho\lambda}{4} \ .
\end{eqnarray}
Thus, the result is consistent with our previous analysis.

Although the anisotropy is always small, it persists during inflation.
Clearly these exact solutions give rise to counter examples to the cosmic no-hair 
conjecture. We should note that the cosmological constant is assumed in
the cosmic no-hair theorem presented by Wald~\cite{Wald:1983ky}. 
In the case of isolated vacuum energy, the inflaton can mimic the 
cosmological constant
~\cite{Kitada:1991ih,Kitada:1992uh,Aguirregabiria:1993pm}. 
However, in the presence of the non-trivial coupling between the inflaton
and the vector field, the cosmic no-hair theorem cannot
be applicable anymore.

%===============================================================%
%************************ SECTION III ***************************%
%===============================================================%
\section{Stability of Anisotropic Inflation}
\label{dynamical}

In the previous section, we found isotropic and anisotropic power-law solutions.
In this section, we will investigate the phase space structure.
Then, we will see which one is dynamically selected.

Let us use e-folding number as a time coordinate $d\alpha = \dot{\alpha} dt$.
It is convenient to define dimensionless variables
\begin{eqnarray}
  X = \frac{\dot{\sigma}}{\dot{\alpha}} \ , \hspace{1cm}
  Y = \frac{1}{M_p} \frac{\dot{\phi}}{\dot{\alpha}} \ , \hspace{1cm}
  Z = \frac{f(\phi)}{M_p} e^{-\alpha +2\sigma} \frac{\dot{v}}{\dot{\alpha}} \ .
\end{eqnarray}
With these definitions, we can write the hamiltonian constraint equation
as
\begin{eqnarray}
  - \frac{1}{M_p^2} \frac{V}{\dot{\alpha}^2}
  = 3(X^2 -1) + \frac{1}{2} Y^2 + \frac{1}{2} Z^2 \ . 
  \label{hamconst}
\end{eqnarray}
Since we are considering a positive potential, we have the inequality
\begin{eqnarray}
3(X^2 -1) + \frac{1}{2} Y^2 + \frac{1}{2} Z^2 < 0 \ .
\label{constraint}
\end{eqnarray}

Using the hamiltonian constraint (\ref{hamconst}), we can eliminate $\phi$
from the equations of motion. 
Thus, the equations of motion can be reduced to the autonomous form:
\begin{eqnarray}
\frac{dX}{d\alpha} &=& \frac{1}{3} Z^2 (X+1) 
+ X\left\{ 3(X^2 -1) + \frac{1}{2} Y^2 \right\} 
\label{eq:X} \,,\\
\frac{dY}{d\alpha} &=& (Y+\lambda) \left\{ 3(X^2 -1) + \frac{1}{2} Y^2 \right\}
+ \frac{1}{3} YZ^2 + \left( \rho + \frac{\lambda}{2} \right)Z^2 
\label{eq:Y} \,,\\
\frac{dZ}{d\alpha} &=& Z \left[ 3(X^2 -1) + \frac{1}{2} Y^2
-\rho Y +1 -2X + \frac{1}{3} Z^2 \right]
\label{eq:Z} \ .
\end{eqnarray}
Therefore, we have 3-dimensional space with a constraint (\ref{constraint}) and the fixed point is defined by $dX/d\alpha=dY/d\alpha=dZ/d\alpha=0$.

First, we seek the isotropic fixed point $X=0$.
From Eq.~(\ref{eq:X}), we see $Z=0$.
The remaining equation (\ref{eq:Y}) yields $Y=-\lambda$ or $Y^2=6$.
The latter solution does not satisfy the constraint (\ref{constraint}). 
Thus, the isotropic fixed point becomes
\begin{eqnarray}
  (X , Y, Z) = ( 0, -\lambda , 0) 
\label{fixedpoint1}\ .
\end{eqnarray} 
This fixed point corresponds to the isotropic power-law solution (\ref{isotropic}).
Indeed, one can check the solution (\ref{iso-power}) leads to the above fixed point.

Apparently, $Z=0$ and $6 X^2 + Y^2 =6$ give a fixed curve. 
However, this contradicts the constraint (\ref{constraint}).

Now, let us find an anisotropic fixed point.
From Eqs.~(\ref{eq:X}) and (\ref{eq:Y}), we have
\begin{eqnarray}
  Y = \left( 3\rho +\frac{\lambda}{2} \right) X - \lambda \ .
  \label{Y0}
\end{eqnarray}
Eq.~(\ref{eq:X}) gives
\begin{eqnarray}
  Z^2 = - \frac{3X}{X+1} \left[ 3(X^2 -1) + \frac{1}{2} Y^2 \right] \ .
  \label{Z2}
\end{eqnarray}
Using the above results in Eq.~(\ref{eq:Z}), we have
\begin{eqnarray}
  \left( X-2 \right) \left[ \left( \lambda^2 + 8 \rho \lambda + 12 \rho^2 +8 \right)X
              - 2 \left( \lambda^2 +2\rho \lambda -4 \right) \right] =0 \ .
\end{eqnarray}

The solution $X=2$ does not make sense because it implies $Z^2 = -18-36 \rho^2 <0$ by Eqs.~(\ref{Y0}) and (\ref{Z2}).
Thus, an anisotropic fixed point is expressed by
\begin{eqnarray}
 X= \frac{2 \left( \lambda^2 +2\rho \lambda -4 \right)}
            {\lambda^2 + 8 \rho \lambda + 12 \rho^2 +8} \ .
\end{eqnarray}
Substituting this result into Eq.~(\ref{Y0}), we obtain 
\begin{eqnarray}
 Y= - \frac{12 \left( \lambda +2\rho  \right)}
            {\lambda^2 + 8 \rho \lambda + 12 \rho^2 +8} \ .
\end{eqnarray}
Eq.~(\ref{Z2}) yields
\begin{eqnarray}
 Z^2 =  \frac{ 18 \left( \lambda^2 +2\rho \lambda -4 \right)
             \left(-\lambda^2 + 4\rho \lambda +12 \rho^2 +8\right) }
            {\left( \lambda^2 + 8 \rho \lambda + 12 \rho^2 +8 \right)^2}  \ .  
\end{eqnarray}
Note that from the last equation, we find $\lambda^2 +2\rho \lambda > 4$ is 
required for this fixed point to exist under inflation $\lambda \ll 1$. 
It is not difficult to confirm that
this fixed point corresponds to the anisotropic power-law solution (\ref{anisotropic}).

Next, we examine the linear stability of fixed points.
The linearized equations for Eqs.~(\ref{eq:X}), (\ref{eq:Y}), (\ref{eq:Z}) 
are given by
\begin{eqnarray}
\frac{d\delta X}{d\alpha} &=& 
  \left( \frac{1}{3}Z^2 + 9 X^2 + \frac{1}{2} Y^2 -3 \right) \delta X
  + XY \delta Y + \frac{2}{3} \left( X+1 \right)Z \delta Z \,,\\ 
\frac{d\delta Y}{d\alpha} &=& 6X\left( Y + \lambda \right) \delta X 
\,,\nonumber\\
& & + \left\{ 3\left( X^2 -1 \right) 
    + \frac{1}{2}Y^2 +Y\left( Y+\lambda \right) 
                + \frac{1}{3} Z^2 \right\} \delta Y 
                + \left( \frac{2}{3}Y +2\rho + \lambda \right) Z\delta Z 
\,,\\
\frac{d\delta Z}{d\alpha} &=& 
2(3X -1)Z\delta X +\left( Y-\rho \right)Z \delta Y 
+\left(3X^2 + \frac{1}{2} Y^2 + Z^2
-2X -\rho Y -2 \right)\delta Z\,.
\end{eqnarray}
In the case of the isotropic fixed point Eq.~(\ref{fixedpoint1}), 
these equations reduce to
\begin{eqnarray}
 \frac{d\delta X}{d\alpha} &=& 
  \left( \frac{1}{2} \lambda^2 -3 \right) \delta X \,,\\ 
\frac{d\delta Y}{d\alpha} &=&  
  \left(  \frac{1}{2} \lambda^2 -3 \right) \delta Y  \,,\\
\frac{d\delta Z}{d\alpha} &=& 
  \left[ \frac{1}{2}\lambda^2 -2 + \rho \lambda \right]\delta Z\,.
\end{eqnarray}
We see that the left hand side of all above equations becomes 
negative when $\lambda^2 +2\rho \lambda < 4$ during inflation
$\lambda \ll 1$, which means 
the isotropic fixed point is an attractor under these conditions
and the isotropic fixed point becomes stable in this parameter
region.
In the opposite case, $\lambda^2 +2\rho \lambda > 4$, 
the fixed point becomes a saddle point and unstable. 

Now we are interested in the fate of trajectories
around the unstable isotropic fixed point.
We will see that those trajectories converge to an anisotropic fixed point. 

Since we are considering the inflationary universe
$\lambda \ll 1$, the condition $\lambda^2 +2\rho \lambda > 4$
implies $\rho \gg 1$. Under these conditions, we can 
approximately write down the linear equations as
\begin{eqnarray}
 \frac{d\delta X}{d\alpha} &=& 
  -3 \delta X \,,\\ 
\frac{d\delta Y}{d\alpha} &=&  
  -3 \delta Y +  \sqrt{6(\lambda ^2 + 2\rho\lambda- 4)} \delta Z \,,\\
\frac{d\delta Z}{d\alpha} &=& 
  - \frac{1}{2}\sqrt{6(\lambda ^2 +2\rho\lambda - 4) } \delta Y \ .
\end{eqnarray}
The stability can be analyzed by setting 
\begin{eqnarray}
  \delta X = e^{\omega \alpha} \delta \tilde{ X} \ , \hspace{1cm}
   \delta Y = e^{\omega \alpha} \delta \tilde{ Y} \ , \hspace{1cm}
    \delta Z = e^{\omega \alpha} \delta \tilde{ Z} \ .
\end{eqnarray}
Then we find the eigenvalues $\omega$ are given by
\begin{eqnarray}
\omega = -3 \ , 
-\frac{3}{2} \pm i \sqrt{3(\lambda ^2 +2\rho\lambda -4)-\frac{9}{4}} \ . 
\end{eqnarray}
As the eigenvalues have negative real part, 
the anisotropic fixed point is stable.
Thus, the end point of trajectories around the unstable isotropic power-law inflation
must be the anisotropic power-law inflation.
\begin{figure}[htbp]
\includegraphics[width=100mm]{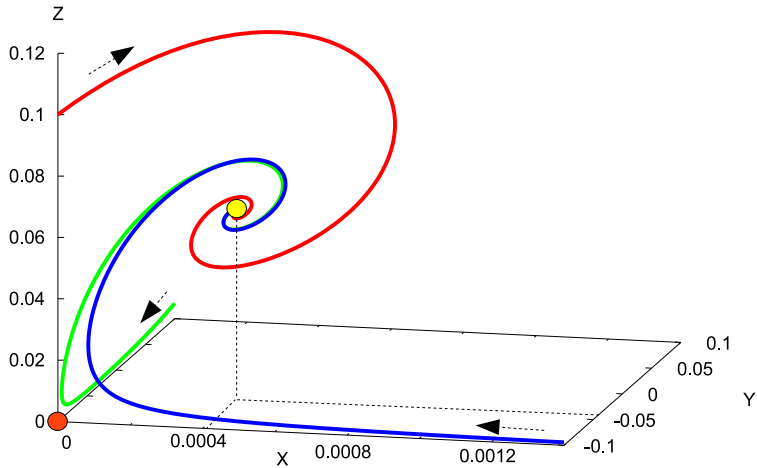}
\caption{The phase flow in $X$-$Y$-$Z$ space is depicted for $\lambda = 0.1, \rho=50$. The yellow and orange circles indicate the anisotropic and isotropic fixed points respectively. The trajectories converge to the anisotropic fixed point.} 
   \label{fg:phase}
\end{figure}
In Fig.\ref{fg:phase}, we depicted the phase flow in $X$-$Y$-$Z$ space for $\lambda = 0.1,\ \rho=50$.
We see that the trajectories converge to the anisotropic fixed point indicated by yellow circle.
The isotropic fixed point indicated by orange circle is a saddle point
 which is an attractor only on $Z=0$ plane.
Thus, the anisotropic power-law  inflation is an attractor
solution for parameters satisfying $\lambda^2 +2\rho \lambda > 4$.

%===============================================================%
%************************ SECTION V ***************************%
%===============================================================%
\section{Conclusion}

We have examined a supergravity model with a gauge kinetic function.
It turned out that exact anisotropic power-law inflationary solutions exist
when both the potential function for an inflaton and the gauge kinetic function
are exponential type. We showed that the degree of the anisotropy is 
proportional to the slow roll parameter. The slow roll parameter depends both on
the potential function and the gauge kinetic function. 
The dynamical system analysis proved that the anisotropic power-law inflation
 is an attractor for a large parameter region $\lambda^2 +2\rho \lambda >4$. Therefore,
the result we have found in this paper indicates that the cosmic no-hair
conjecture should be modified appropriately.

The imprint of the geometrical symmetry breakdown during inflation
can be found in the cosmic microwave background radiation.
We should note that the statistical anisotropy could be large even when
the anisotropy in the expansion is quite small~\cite{Dulaney:2010sq,Gumrukcuoglu:2010yc,Watanabe:2010fh}.

The result in this paper also has implication in cosmological magnetic
 fields~\cite{Kanno:2009ei,Moniz:2010cm}.
In the original paper by Ratra, the same potential and gauge kinetic functions
are assumed~\cite{Ratra:1991bn,Ratra:1991av}. Hence, it is interesting
to incorporate backreaction into his analysis.  

{\sl Note added:} 
Recently, anisotropic inflationary scenarios with other potential functions
have been investigated in \cite{Hassan}. In their paper, a charged scalar
field plays an important role and induces an interesting behavior
around the reheating stage. We thank the authors for a preview of their paper
which is complementary to our work. 

\begin{acknowledgements}
SK is supported in part by grant PHY-0855447 from the National Science 
Foundation. JS is supported by  the
Grant-in-Aid for  Scientific Research Fund of the Ministry of 
Education, Science and Culture of Japan No.22540274, the Grant-in-Aid
for Scientific Research (A) (No. 22244030), the
Grant-in-Aid for  Scientific Research on Innovative Area No.21111006
and the Grant-in-Aid for the Global COE Program 
``The Next Generation of Physics, Spun from Universality and Emergence".
MW is supported by JSPS Grant-in-Aid for Scientific Research No. 22E926.
\end{acknowledgements}

\end{document}